\documentclass[reprint,superscriptaddress,amssymb,amsmath,aps,showpacs,10pt,floatfix,longbibliography,pra]{revtex4-2}

\usepackage{graphicx}  

\usepackage{color}
\usepackage[colorlinks,bookmarks=false,citecolor=darkblue,linkcolor=red,urlcolor=blue]{hyperref}

\definecolor{darkblue}{rgb}{0,0.02,0.45}

\newcommand{\SI}[2]{\ensuremath{#1\,#2}}

\newcommand{\milli}{\ensuremath{\mathrm{m}}}
\newcommand{\kelvin}{\ensuremath{\mathrm{K}}}

\newcommand{\YMGO}{YbMgGaO$_4$}

\newcommand{\Cp}{\ensuremath{C_{\mathrm{p}}}}

\newcommand{\GammaH}{\ensuremath{\Gamma{}_{\mathrm{mag}}}}

\newcommand{\Hpar}{$H\,\|\,c$}
\newcommand{\Hperp}{$H\!\perp\!c$}

\begin{document}

\title{Field evolution of the spin-liquid candidate YbMgGaO$_4$}

\author{Sebastian Bachus}
\email[]{sebastian.bachus@physik.uni-augsburg.de}
\affiliation{Experimental Physics VI, Center for Electronic Correlations and Magnetism, University of Augsburg, 86159 Augsburg, Germany}
\author{Ilia A. Iakovlev}
\affiliation{Experimental Physics VI, Center for Electronic Correlations and Magnetism, University of Augsburg, 86159 Augsburg, Germany}
\affiliation{Theoretical Physics and Applied Mathematics Department, Ural Federal University, Ekaterinburg 620002, Russia}

\author{Yuesheng Li}
%
\author{Andreas W\"o{}rl}
%
\author{Yoshifumi Tokiwa}
\affiliation{Experimental Physics VI, Center for Electronic Correlations and Magnetism, University of Augsburg, 86159 Augsburg, Germany}
\author{Langsheng Ling}
\affiliation{High Magnetic Field Laboratory, Chinese Academy of Sciences, Hefei 230031, China}
\author{Qingming Zhang}
\affiliation{National Laboratory for Condensed Matter Physics and Institute of Physics, Chinese Academy of Sciences, Beijing 100190, China}
\affiliation{School of Physical Science and Technology, Lanzhou University, Lanzhou 730000, China}
\author{Vladimir V. Mazurenko}
\affiliation{Theoretical Physics and Applied Mathematics Department, Ural Federal University, Ekaterinburg 620002, Russia}
\author{Philipp Gegenwart}
\author{Alexander A. Tsirlin}
\email[]{altsirlin@gmail.com}
\affiliation{Experimental Physics VI, Center for Electronic Correlations and Magnetism, University of Augsburg, 86159 Augsburg, Germany}
\date{\today}

\begin{abstract}
We report magnetization, heat capacity, thermal expansion, and magnetostriction measurements down to millikelvin temperatures on the triangular antiferromagnet YbMgGaO$_4$. Our data exclude the formation of the distinct $\frac13$ plateau phase observed in other triangular antiferromagnets, but reveal plateaulike features in second derivatives of the free energy, magnetic susceptibility and specific heat, at $\mu_0H=1.0-2.5$\,T for \Hpar{} and $2-5$\,T for \Hperp{}. Using Monte-Carlo simulations of a realistic spin Hamiltonian, we ascribe these features to nonmonotonic changes in the magnetization and the $\frac12$ plateau that is smeared out by the random distribution of exchange couplings in \YMGO{}.
\end{abstract}

\maketitle

\section{Introduction}

The search for spin-liquid states in triangular antiferromagnets has been intensified by the discovery of Yb-based materials that show three-fold-symmetric arrangement of the magnetic ions, strong magnetic frustration, and persistent spin dynamics down to millikelvin temperatures in zero field~\cite{li2020}. The initial work on YbMgGaO$_4$, where Mg/Ga site disorder causes randomness of exchange couplings~\cite{li2019b}, was recently extended to delafossite materials, such as NaYbX$_2$ (X = O, S, Se), with no visible randomness effects reported to date.

Disorder-free delafossites reveal apparent spin-liquid behavior~\cite{baenitz2018,liu2018} with a broad continuum of magnetic excitations~\cite{ding2019,bordelon2019,ma2020,dai2020} and no signs of spin freezing or magnetic order in zero field~\cite{liu2018,ding2019,sarkar2019}. Magnetic field of about 2\,T applied along the in-plane direction leads to a suppression of the spin-liquid phase~\cite{ranjith2019a,ranjith2019b,ma2020} that evolves into the collinear up-up-down order~\cite{bordelon2019} visible as the $\frac13$ plateau in the magnetization~\cite{ranjith2019b,ma2020}. Details of these transformations and especially the behavior above the $\frac13$ plateau phase remain to be understood~\cite{bordelon2020}, but the overall temperature-field phase diagram is strongly reminiscent of Co-based triangular antiferromagnets that also develop a sequence of field-induced phase transitions with the pronounced $\frac13$ plateau in the magnetization~\cite{shirata2012,susuki2013,koutroulakis2015,quirion2015}. The main difference in this case is the zero-field phase, $120^{\circ}$ order in Co-based materials vs putative spin liquid in Yb-based triangular antiferromagnets~\cite{li2020}.

YbMgGaO$_4$ breaks this analogy, because no field-induced phase transitions were reported in most of the previous studies~\cite{li2015a,paddison2017,zhang2018}, although Steinhardt \textit{et~al.}~\cite{steinhardt2019} detected a crossover by monitoring the shift of the spectral weight from the $M$-point of the Brillouin zone to the $K$-point upon increasing the field. This shift can be paralleled to a magnetization anomaly that would coincide with the anticipated $\frac13$ plateau of a triangular antiferromagnet~\cite{steinhardt2019}.

Here, we report comprehensive field-dependent thermodynamic measurements on \YMGO{} at temperatures down to 40\,mK and in magnetic fields up to 10\,T. Using magnetization and heat-capacity data, we exclude field-induced anomalies in first derivatives of the free energy that would be indicative of a thermodynamic phase transition. We do, however, observe plateaulike features in some of the second derivatives and interpret them as vestiges of field-induced transitions that could occur in \YMGO{} in the absence of randomness effects.

\section{Methods}
\subsection{Experimental Details}
All measurements were performed on \YMGO{} single crystals from Ref.~\cite{li2015b}. No sample dependence was observed in this or any of the previous studies. Excellent crystal quality is confirmed by narrow, resolution-limited peaks in x-ray diffraction and absent paramagnetic impurity contribution probed by electron-spin resonance~\cite{li2015b}. 

Magnetization down to \SI{40}{\milli\kelvin} and in fields up to 10\,T was measured in a dilution refrigerator with a capacitive method using Faraday force magnetometer~\cite{sakakibara1994}. Additionally, measurements up to 7\,T and down to 0.5\,K were performed in a Quantum Design magnetic property measurement system (MPMS) using the $^3$He insert. These data were used to scale the magnetization measured in the dilution refrigerator. In both methods, magnetization is probed directly, and absolute values of the magnetic moment are obtained, unlike in Ref.~\onlinecite{steinhardt2019} where changes in the magnetization are monitored indirectly by a shift in the resonance frequency of the tunnel diode oscillator.  

Heat capacity was measured in the dilution refrigerator down to 200\,mK and up to 7\,T using quasiadiabatic pulse method. The same setup was used for measuring the magnetic Gr\"uneisen parameter $\Gamma_{\rm mag}$~\cite{tokiwa2011}. A weak oscillating magnetic field was superimposed on the main magnetic field, and temperature oscillations due to magnetocaloric effect were detected. They were further used to calculate $\Gamma_{\rm mag}=\left.\frac{1}{T}\frac{\partial{}T}{\partial{}H}\right|_S$.

In-plane thermal expansion and magnetostriction ($\Delta L\!\perp\! c$) were measured in a dilution refrigerator by use of a capacitive dilatometer \cite{Kuechler2012} in fields ($H\perp c$) up to 10\,T and down to temperatures of 100\,mK. 

\subsection{Model and numerical simulations}
Magnetic interactions in \YMGO{} are described by an anisotropic spin Hamiltonian on the triangular lattice~\cite{zhu2018,maksimov2019},

\begin{equation}
 \mathcal H_{\rm exch}= \sum_m\left[\mathcal H_m^{\rm XXZ}+\mathcal H_m^{\pm\pm}+\mathcal H_m^{z\pm}\right],
\label{eq:ham}\end{equation}
which includes interactions between nearest neighbors ($m=1$) and second neighbors ($m=2$). The first term,
\begin{equation}
 \mathcal H_m^{\rm XXZ}=J_m\sum_{\langle ij\rangle}(S_i^xS_j^x + S_i^yS_j^y+\Delta S_i^zS_j^z), 
\label{eq:delta}\end{equation}
is the XXZ Hamiltonian with the exchange anisotropy $\Delta$. The second and third terms,
\begin{align}
 \mathcal H_m^{\pm\pm}=\sum_{\langle ij\rangle} 2J_m^{\pm\pm}[(S_i^x &S_j^x-S_i^yS_j^y)\cos\varphi_{\alpha}- \notag\\[-10pt]
 &-(S_i^xS_j^y+S_i^yS_j^x)\sin\varphi_{\alpha}],
\label{eq:pmpm}\end{align}
\begin{align}
 \mathcal H_m^{z\pm}=\sum_{\langle ij\rangle} J_m^{z\pm}[(S_i^y &S_j^z+S_i^zS_j^y)\cos\varphi_{\alpha}- \notag\\[-10pt]
 &-(S_i^xS_j^z+S_i^zS_j^x)\sin\varphi_{\alpha}],
\label{eq:zpm}\end{align}
stand for the additional anisotropies, including off-diagonal, and $\varphi_{\alpha}={0,\pm2\pi/3}$ is the bond-dependent pre-factor. 

Quantum simulations by exact diagonalization for the spin Hamiltonian of Eq.~\eqref{eq:ham} are restricted to very small lattice sizes. Therefore, we resort to the classical spin Hamiltonian:
\begin{equation}\label{Ham_MC}
  \mathcal H =\mathcal H_{\rm exch}- \sum_{i}\textbf{H}{\bf S}_i - \sum_{i\neq j}B(H)_{ij}({\bf S}_i{\bf S}_j)^2,
\end{equation}
where the exchange Hamiltonian from Eq.~\eqref{eq:ham} is augmented by the Zeeman term and by the biquadratic exchange that emulates the effect of quantum fluctuations~\cite{griset2011}. To determine the optimal size of the biquadratic exchange, we adopt the procedure of Ref.~\cite{griset2011} modified as follows:
\begin{equation}\label{bq}
  B(H^\alpha) = J_1^{zz}[0.0536(1-0.03H^\alpha\sqrt{H^\alpha_{\rm sat}-H^\alpha})]n,
\end{equation}
where $H^\alpha_{\rm sat}$ is the saturation field for the field direction $\alpha$, and $n$ is an integer. The Yb$^{3+}$ $g$-factor ($g_{\perp}$ or $g_{\|}$ depending on the field direction) was added \textit{a posteriori} by rescaling the magnetic field.

Magnetization of the Hamiltonian given by Eq.~\eqref{Ham_MC} was determined by classical Monte Carlo (MC) simulations at temperatures up to $T = 0.15J_1^{zz}$, where $J_1^{zz}=J_1\Delta$, using the Uppsala Atomistic Spin Dynamics (UppASD) package~\cite{skubic2008, eriksson2017}.

Each magnetization curve was calculated using 500 different values of the magnetic field. During the MC simulations, we gradually cool down the system from $T=5.0J_1^{zz}$. Each run is composed of 25 annealing steps with 40 000 MC steps per spin. $\Delta T$ between successive temperature steps is different and set manually. At the final temperature, we perform 100 000 MC steps for thermalization and 150 000 for measurements.

\begin{table}[t!]
\caption{Exchange parameters for the spin Hamiltonian of Eq.~\eqref{Ham_MC}. Models A and B are based on the data from Refs.~\cite{zhang2018,li2015b}, respectively, the latter augmented by the second-neighbor coupling $J_2$.}\label{tab:exc_param}
\begin{ruledtabular}
\begin{tabular}{ccccccc}
        Model & $\Delta$ & $J_1^{\pm\pm}/J_1$ & $J_1^{z\pm}/J_1$ & $J_2/J_1$ & $g_\parallel$ & $g_\perp$ \\\hline
        A & 0.88 & 0.176 &  0.176 & 0.18 & 3.81 & 3.53  \\
        B & 0.54 & 0.086 & 0.02 & 0.18  & 3.72 & 3.06 \\
\end{tabular}
\end{ruledtabular}
\end{table}

The parametrization of $\mathcal H_{\rm exch}$ has been a matter of significant debate~\cite{li2020}. Here, we use two sets of exchange parameters (Table~\ref{tab:exc_param}). Model B based on the Curie-Weiss temperatures and electron spin resonance data~\cite{li2015b} was initially formulated for purely nearest-neighbor (NN) spin Hamiltonian. We augment it by adding the second-neighbor coupling $J_2$ compatible with the neutron scattering data~\cite{paddison2017,zhang2018}. Model A is based on fitting magnetic excitations determined by terrahertz and neutron spectroscopies as a function of field~\cite{zhang2018}. Two of the parameters, $J_1^{\pm\pm}/J_1=0.4(3)$ and $J_1^{z\pm}/J_1=0.6(6)$, are determined with a very high uncertainty, leaving a rather vague definition of the parameter space. We tested several values within the range allowed by the aforementioned experiments, and chose the $J_1^{\pm\pm}/J_1$ and $J_1^{z\pm}/J_1$ values that yield best agreement with the experimental magnetization curves. These values are given in Table~\ref{tab:exc_param} for the parameter set A.

The main difference between the models A and B lies in the extent of the XXZ anisotropy $\Delta$. Both parameter sets place \YMGO{} into the region of a stripe phase in the general phase diagram of triangular antiferromagnets~\cite{zhu2018}, although experimentally \YMGO{} strongly resembles a spin liquid and clearly lacks any magnetic order. Different scenarios of the spin-liquid formation were discussed in the literature~\cite{li2020}. For example, Zhu \textit{et al.}~\cite{zhu2017} argued that structural randomness leads to random directions of spin stripes, and this stripe liquid mimics a spin liquid. On the other hand, Rau and Gingras~\cite{rau2018} suggested that structural randomness will not affect the position of the material on the phase diagram and will only lead to a variation of $J_1$ across the crystal, the scenario reminiscent of a valence-bond solid with quenched disorder~\cite{kimchi2018} that was used to interpret low-energy excitations of YbMgGaO4~\cite{li2019a}. We, therefore, employ the randomness scenario proposed by Rau and Gingras~\cite{rau2018} in our simulations, as further explained in Sec.~\ref{modeling_susceptibility}.


\section{Results}

\subsection{Magnetization and Susceptibility}
\begin{figure*}
\includegraphics[width=0.74\textwidth]{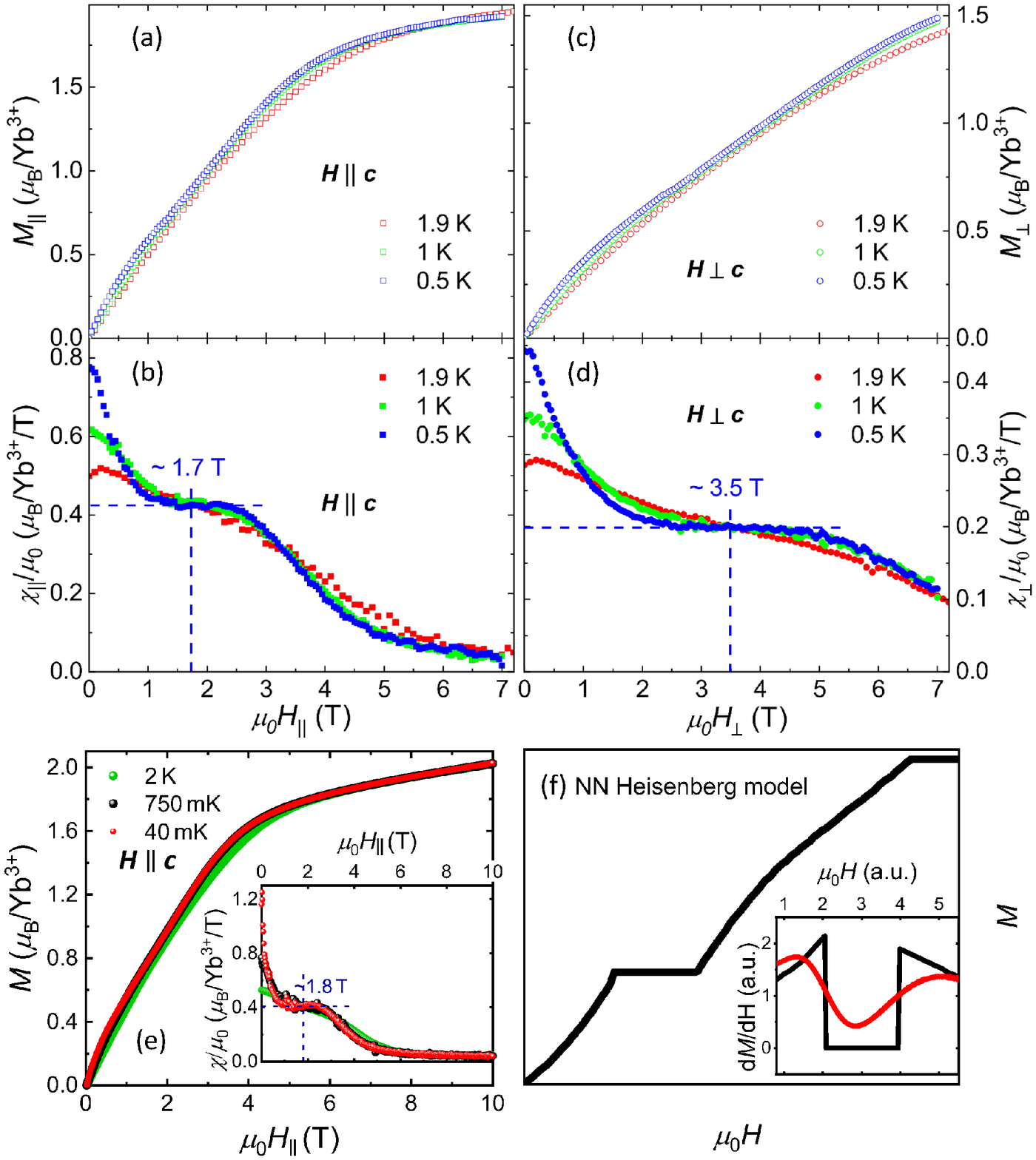}%
\caption{(a,b) Field-dependent magnetization of \YMGO{} measured with MPMS for \Hpar{} and \Hperp{}, respectively. (c,d) Associated magnetic susceptibility $\chi=dM/dH$. (e) Field-dependent magnetization measured for \Hpar{} in the dilution fridge up to 10\,T and down to 40\,mK; the inset shows $\chi(H)$ in good agreement with the MPMS data from panel (c). (f) Schematic picture of field-dependent magnetization for the NN triangular Heisenberg antiferromagnet with $\Delta=1$ and $J_1^{\pm\pm}=J_1^{z\pm}=J_2=0$, as taken from Ref.~[\onlinecite{chubokov1991}]. The $\frac13$ plateau is clearly visible and leads to the rectangular dip in $\chi(H)$ (black line in the inset) that transforms into a minimum when broadening is introduced (red line).
\label{fig.Susceptibility_MPMPS_MK}
}

\end{figure*}

Figure~\ref{fig.Susceptibility_MPMPS_MK} shows field dependence of the magnetization for both directions of the applied field. The saturation magnetization $M_{\rm sat}$ calculated from the experimental $g$-factors (Table~\ref{tab:exc_param}, model B) is about 1.86\,$\mu_B$/Yb$^{3+}$ for \Hpar{} and 1.53\,$\mu_B$/Yb$^{3+}$ for \Hperp{}. These values are reached at the saturation fields of $H_{\rm sat}^{\|}\simeq 5$\,T and $H_{\rm sat}^{\perp}\simeq 7$\,T, respectively. The magnetization increases linearly above $H_{\rm sat}^{\|}$ because of the sizable van Vleck term caused by the higher-lying crystal-field levels of Yb$^{3+}$. The higher value of the saturation field for \Hperp{} reflects easy-plane anisotropy of the exchange couplings. To a first approximation, $\Delta\simeq H_{\rm sat}^{\|}/H_{\rm sat}^{\perp}\simeq 0.71$ is in between the estimates of the models A and B in Table~\ref{tab:exc_param}.

Even at 0.5\,K, the magnetization curves do not show any clear features that could be unambiguously identified as the \mbox{$\frac13$ plateau}, although there is a clear nonlinearity around $M_{\rm sat}/3$ for both field directions  [Figs.~\ref{fig.Susceptibility_MPMPS_MK}(a) and \ref{fig.Susceptibility_MPMPS_MK}(b)]. This effect is better visible in the magnetic susceptibility $\chi=dM/dH$, where a plateau is observed centered around 1.7\,T for \Hpar{} and 3.5\,T for \Hperp{} [Figs.~\ref{fig.Susceptibility_MPMPS_MK}(c) and \ref{fig.Susceptibility_MPMPS_MK}(d)].

To detect these features more clearly, we reduced the measurement temperature to 40\,mK. This was only possible for \Hpar{}, because in the \Hperp{} configuration a strong torque, presumably caused by sample misalignment, prevented reliable measurements with the Faraday magnetometer. Nevertheless, already the \Hpar{} data suggest that reducing the temperature has no visible effect on the non-linearity of $M(H)$ and plateau feature of $\chi(H)$ [Fig.~\ref{fig.Susceptibility_MPMPS_MK}(e)], and they obviously deviate from the prominent \mbox{$\frac13$ plateau} expected, for example, in a NN triangular Heisenberg antiferromagnet [Fig.~\ref{fig.Susceptibility_MPMPS_MK}(f)]. Furthermore, the bend due to saturation remains very broad, much broader than expected at 40\,mK where thermal fluctuations are mostly suppressed. This broadening is an intrinsic effect that reflects a distribution of exchange couplings in \YMGO{} caused by the structural randomness.

Our data are overall in agreement with the indirect magnetization measurement by Steinhardt \textit{et~al.}~\cite{steinhardt2019}, who also observed saturation at $4-5$\,T for \Hpar{} and $7-8$\,T for \Hperp{}. The field-induced crossover proposed in their work seems to coincide with the plateau features of $\chi(H)$. Our direct magnetization measurement clearly excludes a thermodynamic phase transition around this field, because no anomalies are observed in $M(H)$. Both first and second derivatives of the free energy evolve continuously.

\subsection{Modeling of the susceptibility}\label{modeling_susceptibility}
Randomness of exchange couplings broadens all features in the magnetization curves, so it would be natural to interpret the effects observed in our magnetization data as partially smeared out signatures of the \mbox{$\frac13$-plateau}. Two aspects of the data speak against this interpretation, though. First, the $\frac13$-plateau is typically observed for \Hperp{} and not for \Hpar{}, as in Ba$_3$CoSb$_2$O$_9$ ($\Delta=0.85-0.95$)~\cite{susuki2013,quirion2015,koutroulakis2015} and NaYbSe$_2$ ($\Delta=0.49$)~\cite{ranjith2019b}. \YMGO{} with an intermediate value of $\Delta$ shows a quite different behavior, because both field directions lead to very similar signatures in $M(H)$ and $\chi(H)$. These signatures are shifted to higher fields for \Hperp{} because of the stronger in-plane spin components, similar to the anisotropy of the saturation field.

Second, $\chi(H)$ is expected to show a dip, becoming zero at the $\frac13$ plateau while being positive both below and above the plateau [Fig.~\ref{fig.Susceptibility_MPMPS_MK}(f)]. Broadening transforms this dip into a shallow minimum, but it does not lead to a plateau feature with the overall downward trend in $\chi(H)$. We thus conjecture that the features shown in Fig.~\ref{fig.Susceptibility_MPMPS_MK} may have a different origin, and attempt to reproduce them in numerical simulations for a realistic spin Hamiltonian of \YMGO{}.

\begin{figure*}
\includegraphics[width=\textwidth]{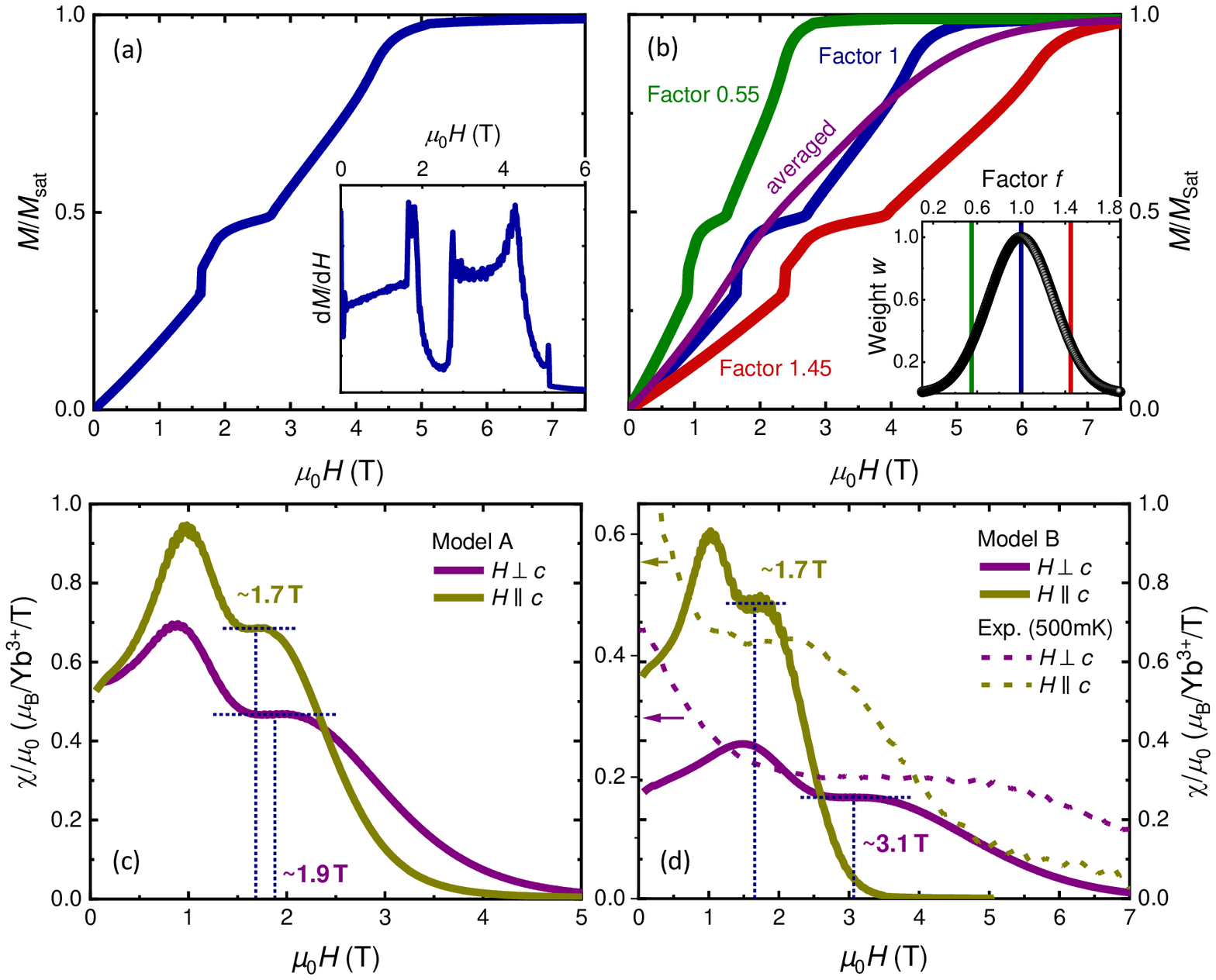}
\caption{(a) The $M(H)$ and $\chi(H)$ curves obtained for model B with $H\perp{}c$, $J_{\rm avg}=2.26$\,K, and $T=0.15J_1^{zz}$ without averaging over different values of $J_1$. No plateau feature comparable to the experimental data is visible in $\chi(H)$. (b) Averaging of the $M(H)$ curve to simulate the exchange randomness. The curve with $f=1$ equals the original curve shown in (a), whereas $f=0.55$ and $f=1.45$ represent two values of $J_1$ which contribute with a smaller weight $w$ to the averaged curve following Eq.~(\ref{eq.avg}). The inset shows a Gaussian distribution of the $f$ values together with the respective weight $w$. The resulting averaged $M(H)$ curve is shown in purple.
(c) Calculated $\chi(H)$ for model A using $J_{\rm avg}=1.63$\,K with $2\sigma=0.53\,J_{\rm avg}$, $n=5$, $T=0.15J_1^{zz}$ for \Hpar{} and $2\sigma=0.61\,J_{\rm avg}$, $n=2$, $T=0.15J_1^{zz}$ for \Hperp{}. (d) Calculated $\chi(H)$ for model B using $J_{\rm avg}=2.26$\,K with $2\sigma=0.46\,J_{\rm avg}$, $n=6$, $T=0.05J_1^{zz}$ for \Hpar{} and $2\sigma=0.6\,J_{\rm avg}$, $n=5$, $T=0.15J_1^{zz}$ for \Hperp{}. This parameter set yields best agreement with the experimental behavior shown by the dotted lines for the experimental data measured at 500\,mK.
\label{fig.Expl_Avg_and_Calc_Susc}}
\end{figure*}

MC calculations for the models A and B from Table~\ref{tab:exc_param} result in magnetization curves with a steplike feature that marks the onset of the $\frac12$ plateau, which is exemplified in Fig.~\ref{fig.Expl_Avg_and_Calc_Susc}(a) for model B, $H\perp c$. Similar behavior is obtained for both field directions within model A. The exception is model B, $H\parallel c$, where only the step was observed, but no plateau was visible. The $\frac12$ plateau is indeed expected for the parameter range of \YMGO{} with $J_2/J_1=0.18$. Quantum simulations for triangular antiferromagnets with large $J_2$ also yield the $\frac12$ plateau and assign it to the up-up-up-down magnetic order stabilized for $J_2/J_1>0.125$~\cite{ye2017}. 

We now compare these results to the experimental data by introducing exchange randomness that stems from the random distribution of unequally charged Mg$^{2+}$ and Ga$^{3+}$ ions in the structure. These nonmagnetic ions impose different electric fields on the magnetic Yb$^{3+}$ ions, affect their crystal-field levels, and influence ground-state wave functions that, in turn, modify exchange couplings in the system. Li \textit{et al.}~\cite{li2017} described these intricate disorder effects using several structural models that reproduce experimental crystal-field excitations probed by neutron spectroscopy. Consequently, Rau and Gingras~\cite{rau2018} used the same structural models to estimate exchange parameters, and showed that only the absolute value of $J_1$ varies throughout the crystal, whereas $\Delta$, $J_1^{\pm\pm}/J_1$, and $J_1^{z\pm}/J_1$ are nearly unchanged. We follow this assumption and employ a Gaussian distribution of $J_1$ in our simulations, while keeping fixed values of $\Delta$ and other exchange parameters given in Table~\ref{tab:exc_param}. We also assume a similar Gaussian distribution of $J_2$ with the constant $J_2/J_1$ ratio. 

Pristine magnetization curve $M(H)_{\rm prist}$ is calculated for $J_{\rm avg}$, an average value of $J_1$ in the crystal, and the field is consequently rescaled by the factor $f=J_1/J_{\rm avg}$ to reflect the change of the energy of exchange couplings of $J_1$. The step and plateau in $M(H)$ are shifted accordingly toward higher or lower fields [Fig.~\ref{fig.Expl_Avg_and_Calc_Susc}(b)]. For each curve, a weighting factor $w$ is assigned using the Gaussian distribution [inset of Fig.~\ref{fig.Expl_Avg_and_Calc_Susc}(b)]. The distribution is truncated at factors where $w=0.01$, which in this example occurs for $J_{1,min/max}=0.1$ and $1.9\,J_{\rm avg}$. It corresponds to a half-width of about 0.7 and a standard deviation of $\sigma \sim 30\%$ of $J_{\rm avg}$. Thus, the interval of $J_{\rm avg}\pm \sigma$ that carries most of the weight corresponds to about 60\% variation in the size of $J_1$ due to structural randomness, in agreement with the estimates of Ref.~\onlinecite{rau2018}. Similar values have been used for the other curves in Figs.~\ref{fig.Expl_Avg_and_Calc_Susc}(c) and \ref{fig.Expl_Avg_and_Calc_Susc}(d), and all parameter are summarized in Table~\ref{tab:avg_gauss_parameter}. 
\begin{table}[t!]
\caption{Summary of the parameter used for the averaging procedure shown in Figs.~\ref{fig.Expl_Avg_and_Calc_Susc}(c) and \ref{fig.Expl_Avg_and_Calc_Susc}(d) and explained in the text. $\sigma $ is defined as the standard deviation. The interval $\pm \sigma$ around $J_{\rm avg}$ includes $\sim 68\%$ of the $J_1$ values, thus carrying most of the total weight.}\label{tab:avg_gauss_parameter}
\begin{ruledtabular}
\begin{tabular}{cccc}
        Model & FWHM (\% of $J_{\rm avg}$)  & $J_{1,min/max}$ & $2\sigma$ (\% of $J_{\rm avg}$) \\
        \hline
        A, $H\parallel{}c$ & 62 & 0.2/1.8\,$J_{\rm avg}$&  53   \\
        A, $H\perp{}c$ & 71 & 0.08/1.92\,$J_{\rm avg}$&  61   \\
        B, $H\parallel{}c$ & 54 & 0.3/1.7\,$J_{\rm avg}$& 46  \\
        B, $H\perp{}c$ & 70 & 0.1/1.9\,$J_{\rm avg}$& 60  \\
\end{tabular}
\end{ruledtabular}
\end{table} 

Averaged magnetization is obtained by adding all factorized curves as follows:
\begin{equation}
M(H)_{\rm avg}=\frac{\sum_{i=0}^{n}w_{i}f_{i}M(H)_{\rm prist}}{\sum_{i=0}^{n}w_i},
\label{eq.avg}
\end{equation}
with 201 $f_i$ values, exemplarily shown again for model B, $H\perp c$, in Fig.~\ref{fig.Expl_Avg_and_Calc_Susc}(b). Averaged curves of $\chi(H)$ for models A and B are shown in Figs.~\ref{fig.Expl_Avg_and_Calc_Susc}(c) and \ref{fig.Expl_Avg_and_Calc_Susc}(d) and, besides predicting the correct order of magnitude, strongly resemble the experimental $\chi(H)$. For model A the plateau in $\chi(H)$ is obtained around the same field for both field directions, whereas for model B the plateau is observed at about twice higher field for \Hperp{} than for \Hpar{} in perfect agreement with the experiment. Experimental positions of the plateaus are reproduced using $J_{\rm avg}=2.26$\,K to be compared with the earlier estimates of 1.8\,K~\cite{li2015b} and 2.0\,K~\cite{zhang2018} that, however, did not take into account exchange randomness.

Several discrepancies should be mentioned as well. Even in model B, the $\chi (H)$ plateaus in the simulated curves are more narrow than in the experiment. This may indicate an approximate nature of our randomness scenario. Moreover, experimental magnetization curves reveal a downward curvature below 1\,T, whereas simulated curves (and their average) are linear in this field range. This difference may arise from quantum effects that, for example, cause nonlinear field-dependent magnetization of frustrated square-lattice antiferromagnets~\cite{thalmeier2008}, but were neglected in our classical MC simulations.

We have also checked how different exchange parameters affect the shape of $\chi(H)$. A small variation of $J_1^{\pm\pm}$ and $J_1^{z\pm}$ does not change the plateaus in $\chi(H)$ or their positions. However, if one of these parameters exceeds 40\% of $J_1^{zz}$, the plateau, as well as the steplike feature in $M(H)$, vanish. This may be related to the fact that off-diagonal terms destabilize the collinear up-up-up-down order forming around $M_{\rm sat}/2$~\cite{ye2017}. Changing $\Delta$ has an immediate effect on the susceptibility and magnetization, as one can see from Figs.~\ref{fig.Expl_Avg_and_Calc_Susc}(c) and \ref{fig.Expl_Avg_and_Calc_Susc}(d). Model B with the smaller $\Delta$ leads to a much better match with the experiment, suggesting a sizable XXZ exchange anisotropy in \YMGO{}. On the other hand, we checked that increasing the anisotropy even further (i.e., reducing $\Delta$ below 0.54) will suppress the features for \Hpar{}. This gives the lower bound of $\Delta\simeq 0.5$ for the \YMGO{} anisotropy.

Finally, broadening of the magnetization curves with the $\frac13$ plateau obtained for $J_2/J_1<0.125$ does not lead to a plateaulike feature in $\chi(H)$ [see Fig.~\ref{fig.Susceptibility_MPMPS_MK}(f)]. This corroborates the sizable $J_2$ in \YMGO{}, as pointed out by neutron and terrahertz spectroscopy~\cite{paddison2017,zhang2018}. 

\subsection{Calorimetry}

\begin{figure*}
\includegraphics[width=0.95\textwidth]{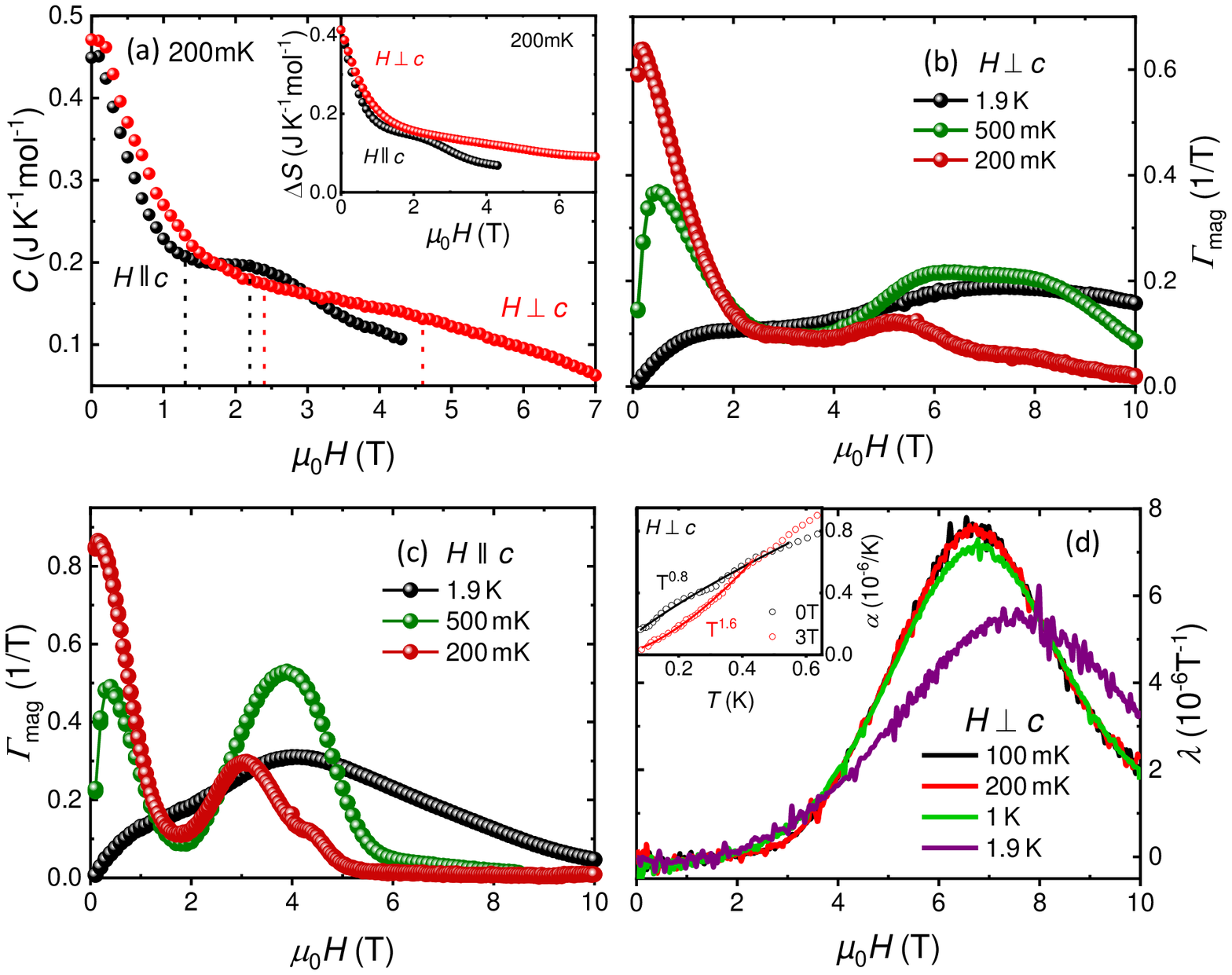}%
\caption{(a) Field dependence of the specific heat $C(H)$ of \YMGO{} measured with \Hpar{} and \Hperp{} at 200\,mK. For both field directions the slope obviously changes in the same field region like the susceptibility, and especially for \Hpar{} a plateau is clearly visible. The dotted lines indicate the position of the plateau in $\chi{}(H)$ for the two field directions, respectively. The inset shows field-dependent magnetic entropy calculated using $C(H)$ and $\GammaH{}(H)$ as described in the text. (b), (c) Field dependence of the magnetic Gr\"uneisen parameter $\Gamma_{\rm mag}$ for different temperatures and both field directions. (d) In-plane magnetostriction $\lambda(H)$ measured at several temperatures for \Hperp{}. The inset shows temperature-dependent linear thermal expansion coefficient $\alpha$ measured at 0 and 3\,T, with the power-law behavior highlighted at low temperatures.
\label{fig.GammaH_HC_Entropy_Magnetostriction}}
\end{figure*}

Calorimetry offers a complementary thermodynamic probe of the field-induced magnetic behavior. In Fig.~\ref{fig.GammaH_HC_Entropy_Magnetostriction}(a), we show specific heat measured as a function of field for both field directions at 200\,mK. Several contributions add up to the total specific heat $C$ probed in our experiment. Electronic contribution is absent in insulating \YMGO{}, whereas lattice contribution is negligibly small below 1\,K~\cite{li2015a}. Therefore, at 200\,mK we are mostly probing magnetic specific heat, probably with a nuclear contribution~\cite{dey2017,ding2019}, but the latter increases with increasing magnetic field, which is not the case in our data. Therefore, we conclude that the signal shown in Fig.~\ref{fig.GammaH_HC_Entropy_Magnetostriction}(a) is dominated by the magnetic specific heat.

Specific heat systematically decreases with field~\cite{li2015a,xu2016,paddison2017}, because low-energy excitations of \YMGO{} are progressively gapped out~\cite{shen2018}. This decrease is nonmonotonic, though. The data for \Hpar{} show a plateau between 1.5 and 2.5\,T in striking resemblance to the plateau in $\chi(H)$ for the same direction of the applied field [Fig.~\ref{fig.Susceptibility_MPMPS_MK}(c)]. Although no clear plateau feature is seen for \Hperp{}, the linear regime between 2.5 and 4\,T is reminiscent of the plateau in $\chi(H)$ in Fig.~\ref{fig.Susceptibility_MPMPS_MK}(d). This confirms that field evolution of \YMGO{} is nonmonotonic. 

Specific heat is proportional to the second derivative of free energy with respect to temperature. In order to reconstruct entropy as the first derivative, we measured the magnetic Gr\"uneisen parameter $\Gamma_{\rm mag}(H)$ which is defined as
\begin{equation}
\Gamma_{\rm mag}{}=\left.\frac{1}{T}\frac{\partial{}T}{\partial{}H}\right|_S.
\end{equation}
The field-dependent entropy $S(H)$ is obtained by integrating the Maxwell relation $\partial{}S/\partial{H}=\partial{}M/\partial{}T$, using $\partial{}M/\partial{}T = -\Gamma_{\rm mag}{}(H)\times{}\Cp{}(H)$~\cite{tokiwa2011,sakai2016}. The resulting $S(H)$ curve was shifted by $0.07\mathrm{R}\ln{2}$ to account for the zero-field entropy at 200\,mK~\cite{li2015a}. 

The overall decrease in the magnetic entropy mirrors the decrease in $C(H)$ due to the low-energy excitations gapped out and the tendency of spins to align parallel to the applied field. A nonmonotonic behavior can be still seen around 2\,T for \Hpar{}, but no features are visible for \Hperp{}. This is similar to the data in Fig.~\ref{fig.Susceptibility_MPMPS_MK}, where only a weak nonlinearity is seen in magnetization as first derivative of the free energy. 

The magnetic Gr\"uneisen parameter itself is a sensitive probe of field-induced phase transitions. Sign change expected for a second-order phase transition is clearly absent across the whole field range in both \Hpar{} and \Hperp{} [Figs.~\ref{fig.GammaH_HC_Entropy_Magnetostriction}(b) and \ref{fig.GammaH_HC_Entropy_Magnetostriction}(c)]. The regions of the $\chi(H)$ and $C(H)$ plateaus correspond to a minimum in $\Gamma_{\rm mag}$, although it does not reach zero, suggesting that $S(H)$ evolves monotonically without going through a maximum, as would be typical for a thermodynamic phase transition. We also note that at 200\,mK $\Gamma_{\rm mag}$ vanishes above 5\,T for \Hpar{} and bends around $7-8$\,T for \Hperp{}. These features coincide with the saturation, which is independently probed via $M(H)$ [Fig.~\ref{fig.Susceptibility_MPMPS_MK}].

\subsection{Dilatometry}
We complete our thermodynamic characterization of \YMGO{} by measuring linear thermal expansion $\alpha=(1/L_0)(dL/dT)$ and magnetostriction $\lambda=(1/L_0)(dL/dH)$. Both parameters are relatively small, on the order of 10$^{-6}$\,K$^{-1}$ and 10$^{-6}$\,T$^{-1}$, respectively, as typical for insulating magnets~\cite{johannsen2005,majumder2018} with only a weak coupling between lattice and spins.

At a constant pressure, specific heat and thermal expansion should have same temperature dependence according to the Gr\"uneisen relation $\alpha(T)=\Gamma\,C(T)$, where $\Gamma$ is the Gr\"uneisen constant. At low temperatures, specific heat of \YMGO{} reveals a peculiar $T^{\gamma}$ power-law behavior with $\gamma\simeq 0.7$ in zero field~\cite{li2015a,xu2016,paddison2017}, reminiscent of $\gamma=\frac23$ in the U(1) quantum spin liquid with spinon excitations. This behavior is indeed reproduced in our zero-field $\alpha(T)$ data that follow the $T^{\gamma}$ power law with $\gamma\simeq 0.8$ below 500\,mK. The power law is thus robust, although its relation to spinons remains debated~\cite{zhu2018,xu2016}, especially in the light of valence-bond models with quenched disorder that can account for this behavior, too~\cite{kimchi2018,li2019a}. At 3\,T, the $\gamma$ value increases to 1.6, indicating gradual opening of a gap in the excitation spectrum. A similar evolution has been seen in the temperature-dependent specific heat of \YMGO{}~\cite{li2015a}.

Positive in-plane magnetostriction indicates that magnetization of \YMGO{} should decrease under uniaxial pressure, according to the Maxwell relation $\lambda V=-(dM/dP)_{p\rightarrow{}0,T,H}$, where $V$ is the molar volume~\cite{stingl2010}. This observation is compatible with $J_1$ increasing under hydrostatic pressure as a result of shortened Yb--O distance and decreased Yb--O--Yb bridging angle~\cite{majumder2020}, because similar structural changes in the YbO$_2$ layer are expected under uniaxial pressure in the $ab$ plane. No thermodynamic anomalies are observed across the whole field range of our study, whereas the maximum around 7\,T can be ascribed to the saturation of the magnetization for \Hperp{}. The broadening of this maximum and the sizable $\lambda$ observed even in 10\,T are compatible with the strong exchange randomness. Below the maximum, $\lambda(H)$ evolves monotonically and does not show plateau-like features observed in $\chi(H)$ and $C(H)$. They are probably fully smeared out by the randomness.

\section{Discussion and Summary}

Triangular antiferromagnets show some of the most intricate field-induced magnetically ordered states~\cite{starykh2015}. Recent experiments suggest that even spin-liquid candidates, which do not order in zero field, are prone to field-induced magnetic order with the formation of the up-up-down collinear phase ($\frac13$ plateau) and possibly other types of magnetic structures~\cite{bordelon2019,bordelon2020}. \YMGO{} stands as an exception. Its field evolution probed by our thermodynamic measurements at millikelvin temperatures does not reveal any anomalies, suggesting that spin dynamics previously confirmed in zero field~\cite{li2016,ding2020} may persist across the whole field range until spins become fully polarized.

Structural randomness and eventual randomness of exchange couplings are the most likely origins of this unusual behavior, because in the presence of randomness any field-induced transition broadens into a cross-over and ultimately fades. Indeed, even at temperatures as low as 40\,mK, we observe very broad features at saturation [Fig.~\ref{fig.Susceptibility_MPMPS_MK}]. Our modeling of the magnetization data, as well as the earlier microscopic results~\cite{rau2018}, suggest at least 50\% distribution of $J_{\rm avg}$ that appears to smear out any field-induced transitions in this system. Nevertheless, even in this highly random setting the field evolution remains somewhat nonmonotonic. We detected plateau features in second derivatives of free energy, magnetic susceptibility and specific heat. The positions of the plateaus reflect the XXZ anisotropy of underlying exchange couplings and of the saturation field. 

Intriguingly, our magnetization data can not be reproduced under an assumption that these features in $\chi(H)$ are vestiges of the $\frac13$ magnetization plateau and associated up-up-down order commonly seen in other triangular antiferromagnets, but they can be modeled if magnetization curves with the $\frac12$ plateau (up-up-up-down order) are considered. The $\frac12$ plateau is expected in triangular antiferromagnets with $J_2/J_1>0.125$, the parameter regime compatible with $J_2/J_1\simeq 0.18$ inferred for \YMGO{} from neutron and terrahertz spectroscopy~\cite{zhang2018}. 

Our modeling of the magnetization process is based on the description of \YMGO{} in terms of a microscopic spin Hamiltonian. A concurrent, phenomenological interpretation can be given within the valence-bond scenario developed in Ref.~\onlinecite{kimchi2018} and backed by the experimental study of low-energy excitations~\cite{li2019a}. This scenario interprets the zero-field ground state of \YMGO{} as a mixture of orphan spins and antiferromagnetic dimers (valence bonds) with randomly distributed exchange couplings. The initial reduction in $\chi(H)$, which is the slope of $M(H)$, can be ascribed to the polarization of these orphan spins, while the plateau region of $\chi(H)$ will reflect the field range where orphan spins are fully polarized, and the dimers are gradually transformed from singlets to triplets. A common aspect of both scenarios is that the field range of the plateau reflects the energy scale of exchange couplings for a given spin direction and thus the extent of the XXZ anisotropy. 

In summary, we have shown that no thermodynamic anomalies occur in \YMGO{} across the whole field range up to saturation, although a nonmonotonic behavior with plateaus features in second derivatives of free energy is observed. The nonlinearity of $M(H)$ and the plateaulike evolution of $\chi(H)$ are seen for both field directions and may not be vestiges of the $\frac13$ magnetization plateau. On the other hand, they can be explained by nonmonotonic changes in the magnetization expected at $J_2/J_1>0.125$. The relative positions of the plateau features for \Hpar{} and \Hperp{} indicate the sizable XXZ anisotropy of \YMGO{}.

\begin{acknowledgments}
We would like to thank Yaroslav Kvashnin for technical assistance with the UppASD package. The work in Augsburg was supported by the German Research Foundation (DFG) via the Project No. 107745057 (TRR80) and by the Federal Ministry for Education and Research through the Sofja Kovalevkaya Award of Alexander von Humboldt Foundation (AAT). The work of I.A.I and M.V.V was supported by the Russian Science Foundation, Grant No. 18-12-00185. Q. M. Z. was supported by the National Key Research and Development Program of China (2017YFA0302904 and 2016YFA0300500) and the NSF of China (11774419 and U1932215).
\end{acknowledgments}

%

\end{document}